        \newif\ifdraft \drafttrue  
        \newif\ifpreparepdf \preparepdftrue 
 \draftfalse

        \ifdraft
\documentclass[prl,aps,preprint,showpacs]{revtex4} 
        \else
\documentclass[prl,aps,twocolumn,showpacs,hyperref]{revtex4} 
        \fi

\ifpreparepdf
    \usepackage[dvips]{color} 
    \usepackage[pdfmark,colorlinks]{hyperref} 
\else 
\fi
    \usepackage{graphicx,amssymb,amsbsy}
\ifpreparepdf 

\else  

\fi

        \ifdraft 
  
  \newcommand{\PC}[1]{$\footnotemark\footnotetext{PC: #1}$}
  
  \newcommand{\DL}[1]{$\footnotemark\footnotetext{Domenico: #1}$}
        \else  
  \newcommand{\PC}[1]{}

  \newcommand{\DL}[1]{}
        \fi

\newcommand{\rf}     [1] {~\cite{#1}}
\newcommand{\refref} [1] {ref.~\cite{#1}}

\newcommand{\refeq}  [1] {(\ref{#1})}
\newcommand{\reffig} [1] {figure~\ref{#1}}

\newcommand{\beq}{\begin{equation}}
\newcommand{\continue}{\nonumber \\ }

\newcommand{\eeq}{\end{equation}}
\newcommand{\ee}[1] {\label{#1} \end{equation}}
\newcommand{\bea}{\begin{eqnarray}}
\newcommand{\ceq}{\nonumber \\ & & }
\newcommand{\eea}{\end{eqnarray}}

\newcommand{\etc}{{\em etc.}}       
\newcommand{\ie}{{i.e.}}            

\newcommand{\statesp}{state space}

\newcommand{\dmn}{\ensuremath{\,d}}  

\newcommand{\fd}{spec\-tral det\-er\-min\-ant}

\newcommand{\optPart}{optimal partition}

\newcommand{\Fokker}{Fokker-Planck}

\newcommand{\ExpaEig}{\Lambda}
\newcommand{\cl}[1]{{n_{#1}}}   
\newcommand{\msr}{{\rho}}               
\newcommand{\pS}{{\cal M}}          
\newcommand{\Lnoise}[1]{{\cal L}^{#1}}    
\newcommand{\Lmat}[1]{{{\bf L}_{#1}}}      
\newcommand{\orbitDist}{{z}}     
\newcommand{\Df}[1]{{f'_{#1}}}
\newcommand{\tr}{\mbox{\rm tr}\,}
\newcommand{\cycle}[1]{\ensuremath{\overline{#1}}}

\begin{document}
\title{
How well can one resolve the state space of a chaotic map?
}
\author{Domenico Lippolis and Predrag Cvitanovi\'c}
\affiliation{
                Center for Nonlinear Science, School of Physics,
                Georgia Institute of Technology,
                Atlanta, GA 30332-0430
               }

\date{\today}

        \begin{abstract}
All physical systems are affected by some noise that limits the
resolution that can be attained in partitioning their state
space. For chaotic, locally hyperbolic flows, this resolution
depends on the interplay of the local stretching/contraction
and the smearing due to noise. We propose to determine the
`finest attainable' partition for a given hyperbolic dynamical
system and a given weak additive white noise, by computing the
local eigenfunctions of the {adjoint Fokker-Planck oper\-ator}
along each periodic point, and using overlaps of their widths
as the criterion for an optimal partition. The Fokker-Planck
evolution is then represented by a {finite} transition graph,
whose spectral determinant yields time averages of
dynamical observables. Numerical tests of such `optimal
partition' of a one-dimensional repeller support our
hypothesis.
        \end{abstract}


\maketitle

\noindent
The effect of noise on the behavior of a nonlinear dynamical system
is a fundamental problem in many areas of science%
\rf{vKampen92,LM94,Risken96},
and the interplay of noise and
chaotic dynamics is of particular current interest%
\rf{gasp02,Fogedby05a,Fogedby06a}.

The purpose of this letter is two-fold. First, and conceptually
the most important, we point out an effect of noise
that has not been addressed in literature: weak noise limits the
attainable resolution of the \statesp\ (`phase space') of a
chaotic system. We formulate the `\optPart'
hypothesis whose implementation requires only integration of a
small set of solutions of the deterministic
equations of motion. Second, more technical
point; we show that the \optPart\ hypothesis replaces the
\Fokker\ PDEs by finite, low-dimensional \Fokker\ matrices, whose
eigenvalues give good estimates of long-time observables
(escape rates, Lyapunov exponents, \etc).

A chaotic trajectory explores a strange attractor, and
evaluation of long-time averages requires effective partitioning
of the \statesp\ into smaller regions. The set of unstable
periodic orbits forms a `skeleton' that can be used to
partition the \statesp\ into such smaller regions, each region
a neighborhood of a periodic point\rf{ruelle,inv} (\ie, a point
on a periodic orbit). The number of periodic orbits grows
exponentially with period length, yielding finer and finer
partitions, with the neighborhood of each periodic orbit
shrinking exponentially.

As there is an infinity of periodic orbits, with each neighborhood
shrinking asymptotically to a point, a deterministic chaotic
system can - in principle - be resolved arbitrarily finely.
However, any  physical system suffers background noise, any
numerical prediction suffers computational roundoff noise, and
any set of equations models nature up to a given accuracy,
since degrees of freedom are always neglected. If the noise is
weak, the short-time dynamics is not altered significantly:
short periodic orbits of the deterministic flow still partition
coarsely  the \statesp. Intuitively, the noise smears out the
neighborhood of a periodic point, whose size is now determined
by the interplay between the diffusive spreading
parameterized\rf{VK79,gasp95} by the diffusion
constant $D$,
and its exponentially
shrinking deterministic neighborhood.
As the periods of periodic orbits increase, the
diffusion always wins, and successive refinements of a
deterministic partition of the \statesp\ stop at the finest
attainable partition,  beyond which the diffusive smearing
exceeds the size of any deterministic subpartition. The smearing
width differs from trajectory to trajectory, so there is
no one single time beyond which noise takes over;
rather,
as we shall show here, the \optPart\ has to be computed for a
given dynamical system and given noise.
This effort brings a handsome practical reward:
as the \optPart\ is finite, the \Fokker\ oper\-ator can be
represented by a finite matrix.

While the general idea is intuitive, nonlinear dynamics
interacts with noise in a nonlinear way, and methods for
implementing the \optPart\ for a given noise still need to be
developed.
In this letter we propose a new approach to this
partitioning. We compute the width of the
leading eigenfunction of the linearized adjoint {\Fokker\ oper\-ator}
on each periodic point.
The \optPart\ is then obtained by tracking the
diffusive widths of unstable periodic orbits until they start
to overlap. We describe here the approach as applied to 1\dmn\
expanding maps; higher-dimensional hyperbolic maps and flows
require a separate treatment for contracting directions, a
topic for a future publication\rf{LipCvi07}.

As the simplest application of the method, consider the orbit
$\{\ldots,x_{-1},x_{0},x_1,x_2,\ldots\}$ of a 1\dmn\ map
$x_{n+1}=f(x_n)$, and the associated discrete Langevin
equation\rf{LM94}
\beq
x_{n+1}=f(x_n)+ \xi_n
\,,
\ee{Langevin}
where the $\xi_n$ are independent Gaussian random variables of
mean 0 and variance $2D$
(the method can be applied to continuous
time flows as well, but a 1\dmn\ map suffices to illustrate the
\optPart\ algorithm).
The corresponding \Fokker\ oper\-ator\rf{Risken96},
\beq
\Lnoise{} \circ \msr_{n}({y}) =
\int 
    \frac{dx}{\sqrt{4\pi D}} \,
    e^{-\frac{(y-f(x))^2}{4D}} \msr_{n}({x})
\ee{DL:dscrt_FP}
carries the density of Langevin trajectories $\msr_{n}(x)$
forward in time to $\msr_{n+1}= \Lnoise{}\circ \msr_{n}$. Since
a density concentrated at point $x_n$ is carried into a density
concentrated at $x_{n+1}$, we introduce local coordinate
systems $\orbitDist_a$ centered on the orbit points $x_a$,
together with a notation for the map \refeq{Langevin}, its
derivative, and, by the chain rule, the derivative of the $k$th
iterate $f^k$ evaluated at the point $x_a$,
\bea
x &=& x_a+\orbitDist_a
    \,, \quad
f_a(\orbitDist_a) = f(x_a+\orbitDist_a)
    \continue
\Df{a} &=& f'(x_a)
    \,, \;\;
f_a^k{}' = \Df{a+k-1} \cdots \Df{a+1}\Df{a}
    \,, \;\;
k \geq 2
\,.
\eea
Here $a$ is the label of point $x_a$, and the label $a\!+\!1 $ is a
shorthand for the next point $b$ on the orbit of $x_a$,
$x_b=x_{a+1}=f(x_a)$. For example, a period-3 periodic point
might have label $a=001$, and by $x_{010}=f(x_{001})$ the next
point label is $b=010$.

If the noise is weak, we can approximate (to leading order in
$D$) the \Fokker\ oper\-ator, $\Lnoise{}_a \circ
\msr_{n}(x_{a+1}+\orbitDist_{a+1}) = \int d
\orbitDist_{a}\Lnoise{}_a(\orbitDist_{a+1},\orbitDist_{a})
\msr_{n}(x_{a}+\orbitDist_{a})$, by linearization centered on
$x_a$, the $a$th point  along the orbit,
\bea
\Lnoise{}_a (\orbitDist_{a+1},\orbitDist_{a}) &=&
(4\pi D)^{-1/2} \,
    e^{-\frac{(\orbitDist_{a+1}-\Df{a}\orbitDist_{a})^2}{4D}}
\,.
\label{DL:lin_FP}
\eea
$\Lnoise{}_a$ maps a Gaussian density
$\msr_{n}(x_{a}+\orbitDist_{a}) = c_a
\exp\left\{-{\orbitDist_{a}^2}/{2\sigma_a^2}\right\}$, of
variance $\sigma_a^2$, into a Gaussian density $\msr_{n+1}(x)$
of variance $\sigma_{a+1}^2= (\Df{a}\sigma_a)^2+ 2D$. This
variance is an interplay of the Brownian noise contribution
$2D$ and the nonlinear contracting/amplifying contribution
$(\sigma\Df{})^2$. The diffusive dynamics of a nonlinear system
are thus fundamentally different from Brownian motion, as the
map induces a history dependent effective noise.

In order to determine the smallest noise-resolvable \statesp\
partition along the trajectory of $x_a$, we need to determine
the effect of noise on the points
preceding $x_a$.
This is achieved by
the {\em adjoint \Fokker\ oper\-ator}
\beq
\Lnoise{\dagger} \circ \tilde{\msr}_{n}({x}) =
 \int 
    \frac{dy}{\sqrt{4\pi D}} \,
    e^{- \frac{(y-f(x))^2}{4D}} \tilde{\msr}_{n}({y})
\,,
\ee{DL:dscrt_adj}
which relates a density $\tilde{\msr}_{n}$ concentrated around
$x_a$ to $\tilde{\msr}_{n-1}= \Lnoise{\dagger} \circ
\tilde{\msr}_{n}$, a density concentrated around
the previous point $x_{a-1}$,
the variance transforming as
$(\Df{a-1}\sigma_{a-1})^2 = \sigma_{a}^2+ 2D$. For an unstable
(expanding) map,
these variances shrink.
After $n$
steps the
variance
is given by
{\small
\beq
(f_{a-n}^{n'}\sigma_{a-n})^2=
    \sigma_{a}^2 +  2D (1+(f_{a-1}')^2 +  \cdots
                       + (f_{a-n+1}^{n-1 '})^2)
\,.
\ee{prl:noiseSum}
}

From the dynamical point of view, a good \statesp\ partition
encodes
the recurrent dynamics; here we shall seek a
partition
in terms of neighborhoods of
periodic points\rf{inv,DasBuch} of short periods.
For the linearized $\Lnoise{\dagger}_{a}$ acting on a fixed
point $x_{a}=f(x_{a})$, the $n \to \infty$ sum
\refeq{prl:noiseSum} converges to a Gaussian of variance
\beq
\sigma_{a}^2 = {2D}/{(\ExpaEig_{a}^{2}-1)}
\,,
\ee{DL:1_eig}
where $\ExpaEig_a = f_a'$,
and for a periodic point
$x_{a} \in p$ to a Gaussian of variance
\beq
\sigma_{a}^2 =
\frac{2D}{1-\ExpaEig_p^{-2}}
\left(\frac{1}{(f_{a}')^2}+ \cdots + \frac{1}{\ExpaEig_p^{2}}\right)
\,,
\ee{DL:n_eig}
where $\ExpaEig_p = f_a^{\cl{p}}{}'$ is the Floquet multiplier
(eigenvalue of the Jacobian linearized flow)
of an unstable ($|\ExpaEig_p|> 1$) periodic orbit $p$ of period
$\cl{p}$.
This is the key formula; note that its evaluation requires no
\Fokker\ formalism,
it depends only on the deterministic orbit and its
linear stability.

We can now state the main result of this letter,
\emph{`the best possible of all partitions'} hypothesis,
as an algorithm:
assign to each periodic point $x_{a}$ a neighborhood of finite
width $[x_{a}-\sigma_{a},x_{a}+\sigma_{a}]$. Consider periodic orbits of
increasing period $\cl{p}$, and {\em stop the process of
refining} the \statesp\ partition as soon as the adjacent
neighborhoods overlap.

\begin{figure}[tbp]
\includegraphics[width=0.28\textwidth]{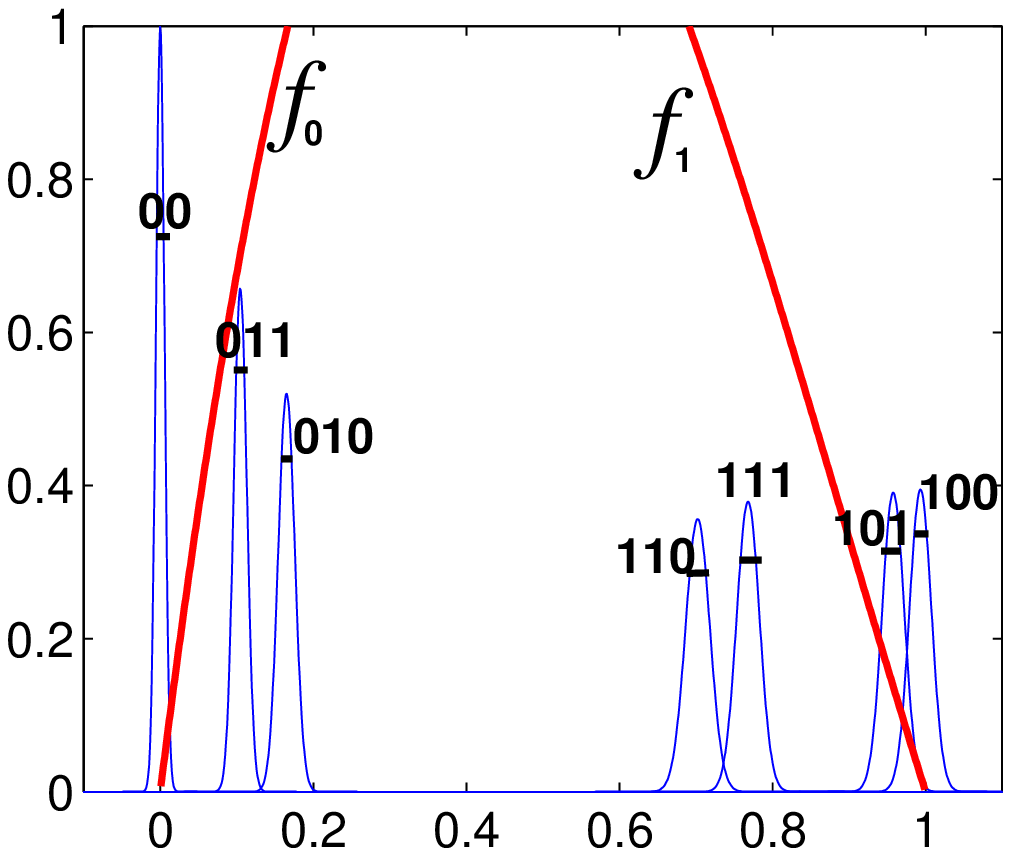}
\caption[]{
$f_0,f_1$: branches of the deterministic map
\refeq{DL:Ulam_rep} for $\ExpaEig_0 = 8$ and $b=0.6$.
The local eigenfunctions $\tilde{\msr}_{a,0}$ with variances
given by \refeq{DL:n_eig} provide a \statesp\ partitioning by
neighborhoods of periodic points of period 3. These are
computed for noise variance  ($D$ = diffusion constant)
$2D=0.002$. The neighborhoods $\pS_{000}$ and $\pS_{001}$
already overlap, so $\pS_{00}$ cannot be resolved further. For
periodic points of period 4, only $\pS_{011}$ can be resolved
further, into $\pS_{0110}$ and $\pS_{0111}$.
}
\label{f:repOverlap}
\end{figure}

As a concrete application to the Langevin map \refeq{Langevin}
consider map\rf{DasBuch}
\beq
f(x) = \ExpaEig_0 x(1-x)(1-b x)
\ee{DL:Ulam_rep}
plotted in~\reffig{f:repOverlap}; this
figure also shows the local eigenfunctions $\tilde{\msr}_{a,0}$
with variances given by \refeq{DL:n_eig}. Each Gaussian is
labeled by the $\{f_0,f_1\}$ branches visitation sequence
of the corresponding
deterministic periodic point (a symbolic dynamics, however, is
not a prerequisite for implementing the method).
\begin{figure}[tbp]
\centering
\includegraphics[width=0.25\textwidth]{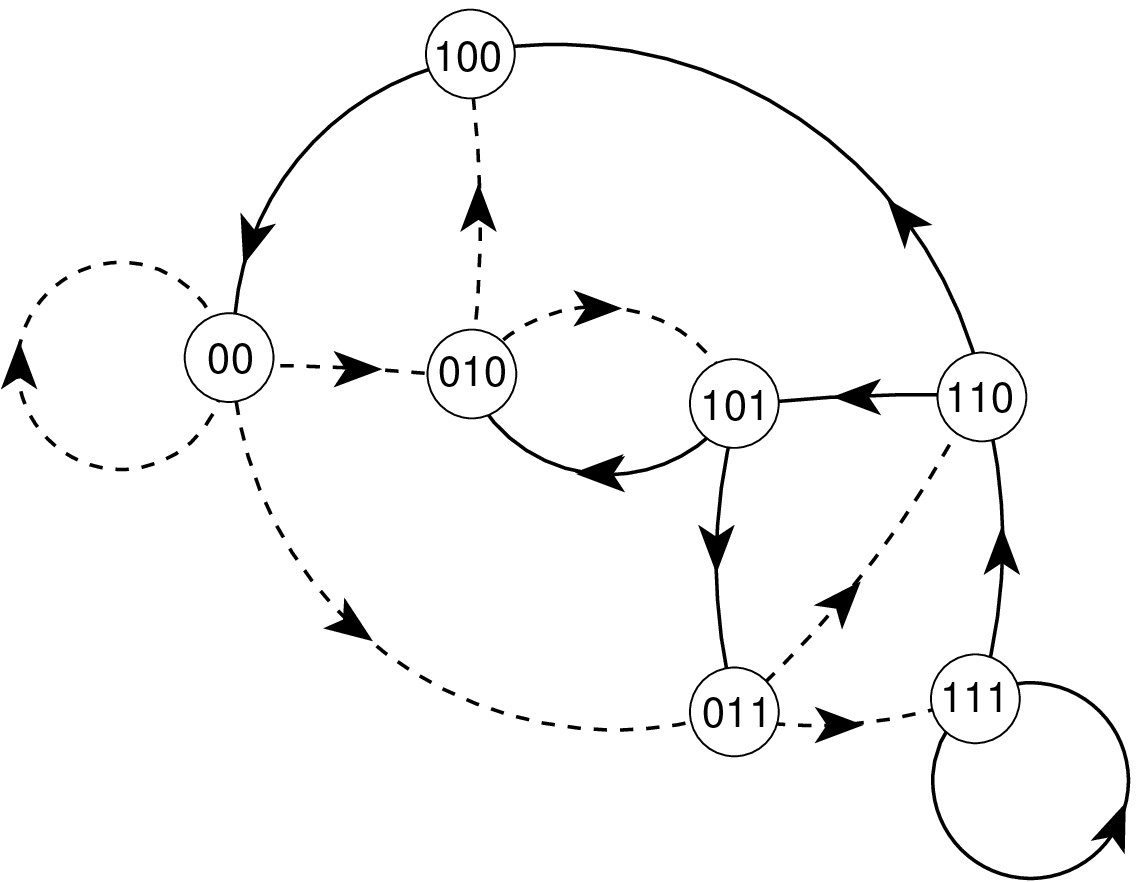}
\caption[]{
Transition graph (graph whose links correspond to the nonzero
elements of a transition matrix $T_{ba}$) describes which
regions $b$ can be reached from the region $a$ in one time
step. The 7 nodes correspond to the 7 regions of the optimal
partition \refeq{TranMatD0.001}. Dotted links correspond to
symbol $0$, and the full ones to 1, indicating that the next
region is reached by the $f_0$, respectively $f_1$ branch of
the map plotted in \reffig{f:repOverlap}.
}\label{f:rep_markov}
\end{figure}
We find that in this case the \statesp\ (the unit interval)
can be resolved into 7 neighborhoods
{\small
\beq
 \{\pS_{00},\pS_{011},\pS_{010},
       \pS_{110},\pS_{111},\pS_{101},\pS_{100}\}
\,.
\label{TranMatD0.001}
\eeq
}
\noindent
Evolution in time maps the \optPart\ interval
$\pS_{011} \to \{\pS_{110},\pS_{111}\}$,
$\pS_{00} \to \{\pS_{00},\pS_{011},\pS_{010}\}$,
\etc,
as compactly summarized
by the transition graph of \reffig{f:rep_markov}.

Next we show that the \optPart\ enables us to
replace \Fokker\ PDEs by finite-dimensional matrices.
The variance \refeq{DL:n_eig} is stationary under the
action of $\Lnoise{\dagger\cl{p}}_{a}$, and the corresponding
Gaussian is thus an eigenfunction.
Indeed, for the linearized flow the entire eigenspectrum is
available analytically, and will be a key ingredient in what
follows. For a periodic point $x_a \in p$, the $\cl{p}$th
iterate $\Lnoise{\cl{p}}_a$ of the linearization
\refeq{DL:lin_FP} is the discrete time version of the
Ornstein-Uhlenbeck process\rf{OrnUhl30}, with left
$\tilde{\msr}_0$, $\tilde{\msr}_1$, $\cdots$, respectively
right ${\msr}_0$, ${\msr}_1$, $\cdots$ mutually orthogonal
eigenfunctions\rf{Risken96} given by
\bea
    \tilde{\msr}_{a,k}(\orbitDist) &=&
    \frac{\beta^{k+1}}{\sqrt{\pi} 2^k k!}
         H_k(\beta \orbitDist) e^{-(\beta \orbitDist)^2}
        \continue
     \msr_{a,k}(\orbitDist) &=& \frac{1}{\beta^k} H_k(\beta \orbitDist)
\,,
\label{DL:Herm}
\eea
where $H_k(x)$ is the $k$th Hermite polynomial,
$1/\beta = \sqrt{2} \sigma_{a}$,
and the $k$th eigenvalue is ${1}/{|\ExpaEig|\ExpaEig^k}$.

Partition \refeq{TranMatD0.001} being
the finest possible partition, the \Fokker\ oper\-ator now acts
as [$7\!\times\!7$] matrix with non-zero $a \to b$ entries
expanded in the Hermite basis,
\bea
[\Lmat{}_{ba}]_{kj} &=&
\left\langle \tilde{\msr}_{b,k} | \Lnoise{} |{\msr}_{a,j} \right\rangle
    \continue
    &=& \int \frac{d\orbitDist_b d\orbitDist_a \,\beta}
              {2^{j+1} j! \pi\sqrt{D}}
    e^{-(\beta \orbitDist_b)^2-\frac{(\orbitDist_b-f_a(\orbitDist_a))^2}{4D}}
    \ceq \qquad \times\,
    H_k(\beta \orbitDist_b) H_j(\beta \orbitDist_a)
\,,
\label{DL:mtx_elem}
\eea
where $1/\beta=\sqrt{2}\sigma_{a}$,
and $\orbitDist_a$ is the deviation from the periodic point $x_a$.
It is the number of resolved
periodic points that determines the dimensionality of the
\Fokker\ matrix.

Periodic orbit theory\rf{DasBuch,PG97} expresses the long-time
dynamical averages, such as Lyapunov exponents, escape rates,
and correlations, in terms of the leading eigenvalues of the
\Fokker\ operator $\Lnoise{}$.
In our
`\optPart' approach, $\Lnoise{}$ is approximated by the
finite-dimensional matrix $\Lmat{}$, and its
eigenvalues are determined from
the zeros of $\det(1-z\Lmat{})$,
expanded as a polynomial in $z$, with
coefficients given by traces of powers of $\Lmat{}$. As the
trace of the $n$th iterate of the \Fokker\ operator
$\Lnoise{n}$ is concentrated on periodic points $f^n(x_a)=x_a$,
we evaluate the contribution of periodic orbit $p$ to $\tr
\Lmat{}^\cl{p}$ by centering $\Lmat{}$ on the periodic orbit,
\beq
t_p = \tr_{p}\, \Lnoise{\cl{p}}
    =
\tr\Lmat{ad}\cdot\cdot\cdot\Lmat{cb}\Lmat{ba}
\,,
\ee{DL:tps}
where $x_a, x_b, \cdots x_d \in p$ are successive periodic
points.
To leading order in the noise variance $2D$, $t_p$ takes the
deterministic value $t_p =1/|\ExpaEig_p-1|$.

We illustrate the method by calculating the escape rate
$\gamma = - \ln z_0$, where
$z_0^{-1}$
is the leading eigenvalue of
\Fokker\ operator $\Lnoise{}$,
for the repeller plotted in~\reffig{f:repOverlap}. The \fd\
can be read off the transition graph of~\reffig{f:rep_markov},
\bea
&& \det(1-z\Lmat{}) =
    1 - (t_0 + t_1)z - (t_{01} - t_0 t_1)\,z^2
    \ceq\quad
      - (t_{001} + t_{011} -  t_{01} t_0 - t_{01} t_1)\,z^3
    - \cdots
    \ceq
    \quad - (t_{0010111} + t_{0011101} - \cdots + t_{001} t_{011} t_1)\,z^7
.
\label{PC:rep_det}
\eea
The polynomial coefficients are given by products of
non-intersecting loops of the transition graph\rf{DasBuch},
with the escape rate
given by the leading root
$z_0^{-1}$
of the polynomial.
Twelve periodic orbits
\cycle{0}, \cycle{1}, \cycle{01}, \cycle{001}, \cycle{011},
\cycle{0011}, \cycle{0111}, \cycle{00111},
\cycle{001101}, \cycle{001011},
\cycle{0010111}, \cycle{0011101}
up to period 7 (out of the 41	
contributing to the noiseless, deterministic cycle expansion
up to cycle period 7) suffice to fully determine
the \fd\ of the \Fokker\ oper\-ator.
%
\begin{figure}[tbp]
\includegraphics[width=0.45\textwidth]{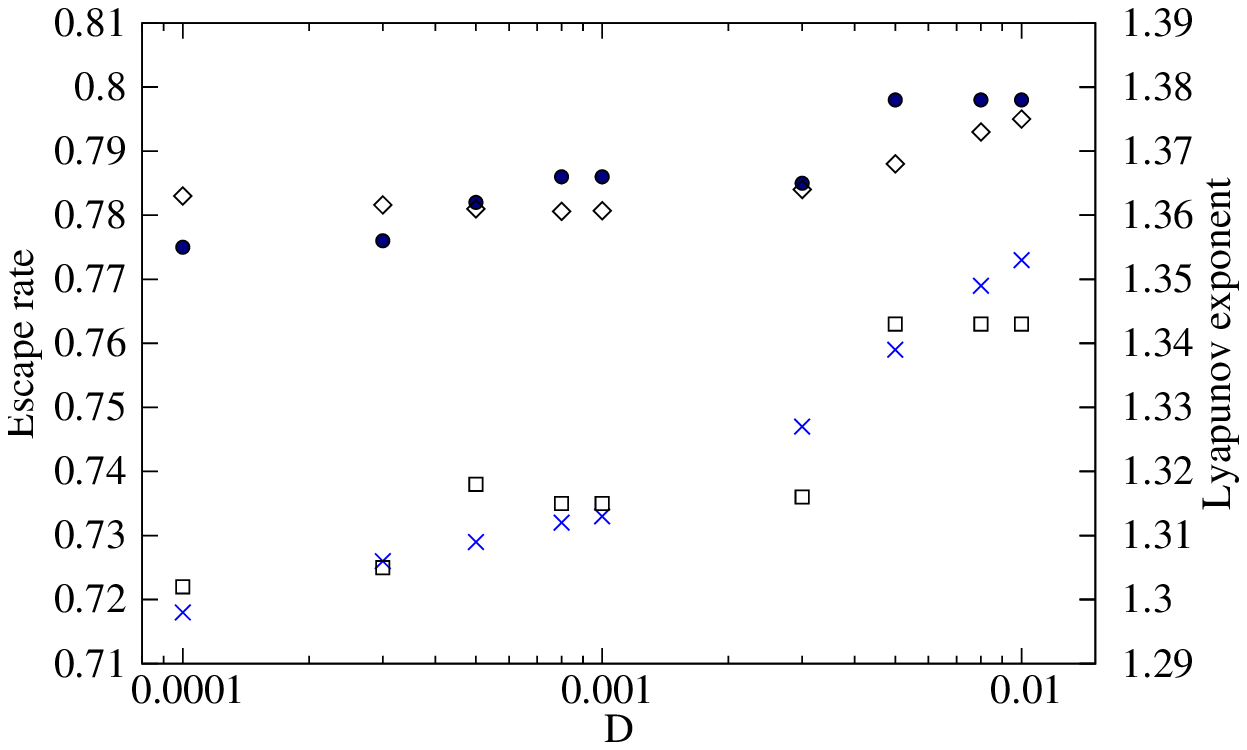}%
\caption[]{
(left scale) the escape rate of the repeller \refeq{DL:Ulam_rep} vs. the noise
strength $D$, calculated using
($\square$) the `\optPart', and ($\color{blue}\times$) a uniform discretization
\refeq{overl-intFP} in $N=128$ intervals;
(right scale) the Lyapunov exponent of the same repeller vs. $D$,
estimated using ($\bullet$) the `\optPart', and ($\diamond$)
the average \refeq{DL:num_liap}.}
\label{f:escRates}
\end{figure}
The `\optPart' estimate of the Lyapunov exponent is given\rf{DasBuch}
by
$\lambda = \left<\ln\,\Lambda\right>/\left<n\right>$,
where the cycle expansion average of an observable $A$
\bea
\left<A\right> &=& A_0t_0+A_1t_1 + (A_{01}t_{01}-(A_0+A_1)t_0t_1) +
\ceq\quad 	
              	   (A_{001}t_{001}-(A_{01}+A_0)t_{01}t_0) + \cdots
\label{DL:average}
\eea
is the finite sum over cycles contributing to \refeq{PC:rep_det},
and $\ln \Lambda_p = \sum\ln|f'(x_a)|$ sum over the points of cycle
$p$ is the cycle Lyapunov exponent.	

Since our `\optPart' algorithm is based on a sharp overlap criterion,
small changes in noise strength $D$ can lead to transition graphs
of different topologies. We assess the accuracy of our finite
\Fokker\ matrix approximations
by discretizing the \Fokker\ operator $\Lnoise{}$
with a piecewise-constant approximation on a uniform mesh
on the unit interval\rf{Ulam60},
\beq
[\Lnoise{}]_{ij} \,=\, \frac{1}{\sqrt{4\pi D}}
    \int_{\pS_i} \frac{dx}{|\pS_i|}   \int_{f^{-1}(\pS_j)} 
    \!\!\!\!\!\!\!\!\!\!\!\!dy \,
    e^{-\frac{1}{4D}(y-f(x))^2}
\,,
\ee{overl-intFP}
where $\pS_i$ is the $i$th interval in equipartition of the
unit interval into $N$ pieces. Empirically,
$N=128$ intervals
suffice to compute the leading eigenvalue of the discretized
$[128 \times 128]$ matrix $[\Lnoise{}]_{ij}$ to
four significant digits.
The Lyapunov exponent is evaluated as the average
\beq
\lambda = \int dx \,e^\gamma \rho(x)\ln|f'(x)|
\ee{DL:num_liap}
where $\rho(x)$ is the leading eigenfunction of \refeq{overl-intFP},
$\gamma$ is the escape rate, and $e^\gamma \rho$
is the normalized repeller measure,
$\int dx  \, e^{\gamma}\rho(x) = 1$.
The numerical results are summarized in \reffig{f:escRates}, with the
estimates of the `\optPart' method within $1\%$ of those
given by the uniform discretization of \Fokker.

In summary, we have presented a new method for partitioning the
\statesp\ of a chaotic repeller in the presence of noise. The
key idea is that the width of the linearized adjoint \Fokker\
operator $\Lnoise{\dagger}_a$ eigenfunction computed on a
periodic point $x_a$ provides the scale beyond which no further
local refinement of \statesp\ is possible. This computation
enables us to systematically determine the {\em \optPart}, of
the finest \statesp\ resolution attainable for a given chaotic
dynamical system and a given noise.
Once the \optPart\ is determined, we use the associated transition
graph to describe the stochastic dynamics by a {\em finite
dimensional} \Fokker\ matrix. While an expansion of the
\Fokker\ oper\-ator about periodic points was already introduced in
\refref{diag_Fred}, the novel aspect of this work is its
representation in terms of the eigenfunctions of the linearized
\Fokker\ oper\-ator~\refeq{DL:lin_FP}, {\em ie.} the Hermite
basis\rf{LipCvi07,gasp95}.

We test our \optPart\ hypothesis by applying it to evaluation
of the escape rate and the Lyapunov exponent of a $1d$ repeller
in presence of additive noise. Numerical tests indicate that,
the `\optPart' method can be as accurate as a $128$-interval
discretization of the \Fokker\ oper\-ator.

The success of the \optPart\ hypothesis in a 1-dimensional
setting is encouraging. However, higher-dimensional hyperbolic
maps and flows, for which an effective \optPart\ algorithm
would be very useful, present new challenges due to  the
subtle interactions between expanding, marginal and
contracting directions.
The nonlinear diffusive effects
(weak stochastic corrections\rf{diag_Fred})
need to be accounted for as well.
These issues will be addressed in a future
publication\rf{LipCvi07}.

\acknowledgements

We are indebted to C.P.~Dettmann, W.H.~Mather, A.~Grigo and
G.~Vattay for many stimulating discussions, and
S.A.~Solla for a critical reading of the manuscript. P.C.
thanks Glen P.~Robinson and NSF grant DMS-0807574 for partial
support.

 \bibliography{../bibtex/lippolis}

\end{document}